\documentclass[a4paper,12pt]{article}
\makeatletter
\def\fmslash{\@ifnextchar[{\fmsl@sh}{\fmsl@sh[0mu]}}
\def\fmsl@sh[#1]#2{%
  \mathchoice
    {\@fmsl@sh\displaystyle{#1}{#2}}%
    {\@fmsl@sh\textstyle{#1}{#2}}%
    {\@fmsl@sh\scriptstyle{#1}{#2}}%
    {\@fmsl@sh\scriptscriptstyle{#1}{#2}}}
\def\@fmsl@sh#1#2#3{\m@th\ooalign{$\hfil#1\mkern#2/\hfil$\crcr$#1#3$}}
\makeatother
\usepackage{epsfig}
\usepackage{epstopdf}
\usepackage{braket}
\usepackage[sumlimits,intlimits,namelimits]{amsmath} 

\usepackage[english]{babel}

\usepackage{graphicx}

\usepackage{geometry}
\numberwithin{equation}{section}
\begin{document}
\begin{titlepage}
\begin{flushright}
SI-HEP-2016-25  \\ 
QFET-2016-14 \\[0.2cm]
\today
\end{flushright}

\vspace{1.2cm}
\begin{center}
{\Large\bf 
Revisiting Uraltsev's BPS limit for Heavy Quarks}
\end{center}

\vspace{0.5cm}
\begin{center}
{\sc Johannes Heinonen} and {\sc Thomas Mannel} \\[0.1cm]
{\sf Theoretische Physik 1,
Universit\"at Siegen, D-57068 Siegen, Germany}
\end{center}

\vspace{0.8cm}
\begin{abstract}
\vspace{0.2cm}\noindent
Motivated by the recent experimental data of the values of the Heavy Quark Expansion parameters, in particular the 
spin-orbit and Darwin terms, we argue that nature actually may be close to a limit of QCD which has been suggested 
by Uraltsev more than ten years ago. Assuming that this limit is not accidental,  we derive the relations among the 
Heavy Quark Expansion parameters that occur up to the order $1/m_b^5$.  
\end{abstract}

\end{titlepage}

\newpage
\pagenumbering{arabic}
\section{Introduction}
The heavy quark expansion (HQE) has turned out to be a very efficient tool in the theoretical description of 
decay processes for bottom hadrons~\cite{Manohar:2000dt}. In particular, for inclusive quantities like lifetimes and $b\to c$ semileptonic 
decays, the HQE has been very successful, including now also higher orders in the $1/m_b$ expansion~\cite{Gambino:2013rza}. 

However, at higher orders in the HQE the number of new, nonperturbative parameters grows dramatically~\cite{Dassinger:2006md,Mannel:2010wj} 
These parameters are given as forward matrix elements of local operators involving an increasing number of 
derivatives.  The leading order is given by dimension-three operators of the form $\bar{Q} \Gamma Q$; these matrix 
elements can all be reduced to the normalized matrix element of the heavy quark current $\bar{Q} \gamma_\mu  Q$, 
up to higher order terms in $1/m_Q$. 

For the dimension-four operators heavy-quark symmetries and the equations of motion ensure that the forward matrix elements of 
all these operators can be expressed
in terms of the matrix elements of higher dimensional operators, so the first nontrivial contributions appear at dimension five, where two independent 
parameters appear. This remains true at dimension six, while at dimension seven already nine independent parameters 
appear. At dimension eight one obtains already eighteen independent matrix elements, and for higher orders one has 
an almost factorial increase of the number of independent parameters. This has lead to the speculation that the HQE actually 
is an asymptotic expansion, very much like the usual perturbative expansion in powers of the strong coupling in 
$\alpha_s$~\cite{Bigi:1997fj}. 

Nevertheless, from the practical side any information on these matrix elements is welcome. While the parameters at dimension 
five and six have been extracted form the data on inclusive semileptonic decays, this becomes impossible for the nine parameters 
at dimension seven, let alone for the eighteen parameters at dimension eight. The only information comes from a naive factorization 
idea suggested in~\cite{Mannel:2010wj}  and elaborated in~\cite{Heinonen:2014dxa}.  

To this end, we revisit a consideration originally 
put forward by Nikolai Uraltsev \cite{Uraltsev:2003ye} in which a relation between the spin-singlet and spin-triplet operator was suggested. 
At the time of publication of \cite{Uraltsev:2003ye}  there was only the observation that the values for the kinetic energy parameter 
$\mu_\pi^2$ is in fact close to the one of the chromomagnetic moment $\mu_G^2$, such that the limit  $\mu_\pi^2 = \mu_G^2$ 
serves as a reasonable starting point. 

Recently the two parameters at dimension six have been extracted from the data, which are the Darwin term $\rho_D^3$ and the 
spin-orbit term $\rho_{LS}^3$. The limit proposed in \cite{Uraltsev:2003ye} implies    $\rho_D^3 + \rho_{LS}^3 = 0$ which is 
actually found in the fit. 

To this end, we take the limit (which we shall call NU limit in what follows) seriously and investigate its implications for the higher-order 
matrix elements, including the dimension-seven and -eight matrix elements. The relations obtained from the NU limit basically relate the 
matrix elements of the spin-singlet wth the spin-triplet operators and thus roughly reduce the number of independent parameters 
by a factor of two.  
 
In the next section we shall discuss the physics reasoning for the NU limit, which we motivate by looking into the phenomenology 
of the lowest-order matrix elements in section~\ref{HOM}.  Based on the NU limit we 
consider the implications for the HQE matrix elements as they appear in the higher orders in the $1/m_b$ expansion .  

\section{A BPS-like limit in Heavy Quark Theory}
In this section we will give a motivation for a specific limit of QCD with heavy quarks, 
which seems to be phenomenologically successful. We emphasize that we do not claim 
to prove this limit, rather we give some phenomenological evidence, and we will try to analyze 
this idea a bit more detailed as it has been done in the original paper \cite{Uraltsev:2003ye} . 

The original idea can be easily understood by looking at a simple system, which is a nonrelativistic 
quark moving in a constant external chromomagnetic field, which we chose along the $z$ direction. It is well known 
that the orbital motion in the $x,y$ plane is quantized, yielding the Landau levels. The Hamiltonian for the 
orbital motion in the $x,y$ plane is constructed form the kinetic momentum with the components  
\begin{equation}
\vec{\Pi} = \vec{p} - g_s \vec{A}  \quad \mbox{with} \quad \vec{A} = \frac{1}{2} \, \vec{r} \times \vec{B}  
\end{equation} 
where $\vec{B}$ is the constant and homogeneous chromomagnetic field with its potential $\vec{A}$.  
The two components of the momentum have c-number commutator relations 
\begin{equation}
\left[ \Pi_x \, , \, \Pi_y \right] = e B \quad \mbox{with} \quad B = | \vec{B} |  \quad \mbox{and} \quad \vec{B} = B \vec{e}_z
\end{equation} 
From this we obtain for the motion in the $x,y$ plane the Hamiltonian for a harmonic oscillator
\begin{equation}
H_o = \frac{1}{2m} \left( \Pi_x^2 + \Pi_y^2 \right) =  \omega \left(a^\dagger a + \frac{1}{2} \right)
\end{equation}  
where $\omega = g_s B / m$ is the cyclotron frequency and 
\begin{equation}
a = \sqrt{\frac{1}{2 g_s B}} \left( \Pi_x  + i  \Pi_y \right)   \, . 
\end{equation} 

Adding now the interaction of the chromomagnetic field with the quark spin with the tree level $g$ factor $g=2$ yields 
\begin{equation}
H = H_o - \frac{g_s}{m} \vec{s} \cdot \vec{B} = \omega \left(a^\dagger a + \frac{1}{2} [1 - \sigma_z] \right)
\end{equation} 
Thus the ground state with $\sigma_z = 1$ has exactly zero energy, while the excited states always appear in 
degenerate pairs for the two spin directions. This toy system is actually a simple and well known expample for a 
supersymmetry in quantum mechanics: One may define a supercharge 
\begin{equation}
S = \sqrt{\frac{1}{2m}} ( \vec{\sigma} \cdot  \vec{\Pi} )   |_{x,y}  = \sqrt{\frac{1}{2m}} \left( \sigma_x \Pi_x + \sigma_x \Pi_x \right) 
\end{equation} 
which is  the ``square root'' of the Hamiltonian $H = S^2$ and which generates a (super)symmetry. 
The ground state (BogomolÕnyi, Prasad, Sommerfield (BPS) state \cite{Prasad:1975kr}) is invariant under this symmetry, corresponding to 
\begin{equation} \label{BPS} 
S | BPS \rangle = \sqrt{\frac{1}{2m}} ( \vec{\sigma} \cdot  \vec{\Pi} )   |_{x,y} \,\, | BPS \rangle = 0 
\end{equation} 
while for the excited states the operator $S$ links the two degenerate states. 

In analogy to this reasoning Uraltsev proposed  in~\cite{Uraltsev:2003ye} that the $B$ meson state may be such a state, such that the operator 
$\vec{\sigma} \cdot \vec{\Pi}$ acting on the state of a $B$ meson is zero, drawing the analogy with 
(\ref{BPS}). Note that for four component spinors we have 
\begin{equation}  \label{nrsB} 
\vec{\sigma} \cdot \vec{\Pi} = \gamma_0 \gamma_5 \vec{\gamma} \cdot \vec{\Pi}  \, . 
\end{equation} 
Phenomenologically this immediately leads to the 
conclusion that in this limit $\mu_\pi^2 = \mu_G^2$; this will be discussed in more detail in section~\ref{HOM}.  

However, this reasoning was in the framework of the simple quantum mechanics of a non-relativistic quark in 
an external chromomagnetic magnetic field,  and it remained unclear how this can be transferred to a quantum field theory 
such as QCD. In particular, there is no evident (approximate) degeneracy between particle states 
which corresponds to the one in the quantum mechanical system. 

For our analysis in QCD the starting point is the Dirac equation for a heavy quark $Q$ with mass $m$, which eventually 
has to become the equation of motion for the field operator $Q$ in the corresponding quantum field theory. Using the usual 
rephasing of the heavy quark field 
\begin{equation}
Q (x) = \exp (-i m (vx) ) \, Q_v (x) 
\end{equation} 
which removes the large part of the heavy quark momentum. Indeed, the QCD covariant derivative 
$D = \partial + i g_s A$ ($i \vec{D} = \vec{\Pi}$) acting on $Q$ 
yields 
$$
i D_\mu Q(x) = (m v_\mu + i D_\mu)  Q_v (x)
$$
which turns the Dirac equation $(i \fmslash{D} - m) Q = 0$ for the heavy quark into
\begin{equation}
[ i \fmslash{D} + (\fmslash{v} - 1) m  ] Q_v = 0 
\end{equation} 
The covariant derivative may be split according to 
$$
i D_\mu  = v_\mu  (i v D) + i D_\mu^\perp  
$$
where (vD) corresponds to the time derivative and $D^\perp$ ($\equiv \vec{D}$) to the spatial components. 
In this way we can write the Dirac equation in hamiltonian form 
\begin{equation}  \label{SG} 
(iv \partial) Q_v = [  (\fmslash{v} - 1) m + g_s (vA) - \fmslash{v} i \fmslash{D}^\perp ] Q_v 
\end{equation} 

On the other hand, in the quantum field theory there is the equation of motion for the field operator 
\begin{equation}
i \partial_\mu Q(x) = [P_\mu \, , \, Q(x) ]  \quad \mbox{or} \quad (m v_\mu + i \partial_\mu)  Q_v (x) = [P_\mu \, , \, Q_v (x) ] 
\end{equation} 
which has to be compatible with (\ref{SG}) 
\begin{equation}
(iv\partial) Q_v =  [Q_v (x)  \, , \, H ] - m Q_v \quad \mbox{with} \quad H = (vP)  \, .
\end{equation}

We may use this operator equation and act on the $0^-$ ground state $ | B (v) \rangle$ of a meson consisting 
of a heavy quark $Q$ and light degrees of freedom. Assuming that  $| B (v) \rangle$ is an eigenstate of the 
full QCD Hamiltonian with eigenvalue $M_B$, we get 
\begin{equation}
(iv\partial) Q_v  | B (v) \rangle =  (M_B - m)  Q_v (x) | B (v) \rangle  - H Q_v (x) | B (v) \rangle 
\end{equation}
as an equation of motion for the state $Q_v (x) | B (v) \rangle $. 

The state $Q_v (x) | B (v) \rangle $ can be interpreted as the state of the light degrees of freedom  inside the 
meson state $| B (v) \rangle $, since $Q_v$ annihilates the heavy quark inside the meson. Thus 
$H Q_v (x) | B (v) \rangle$ corresponds to the energy of the light degrees of freedom; assuming that 
$Q_v (x) | B (v) \rangle $ is also an eigenstate of the Hamiltonian, its must have the eigenvalue 
$\bar\Lambda = M_b - m$, 
resulting in 
\begin{equation}
(iv\partial) Q_v  | B (v) \rangle =  (M_B - m - \bar\Lambda)  Q_v (x) | B (v) \rangle  = 0  \, .
\end{equation}

We may now perform the same steps for the Dirac equation, which in full QCD is the resulting equation of 
motion of the heavy quark field. For simplicity we use the temporal gauge, such that $(ivD) = (iv \partial)$ 
and hence we get 
\begin{equation}
(iv\partial)  Q_v (x) | B (v) \rangle 
=   [  (\fmslash{v} - 1) m - \fmslash{v} i \fmslash{D}^\perp ]  Q_v (x) | B (v) \rangle
\end{equation}   
The first term on the right-hand side projects out the ``large'' components of the spinor field, 
while the second projects out the ``small' components. In order to have the two equations compatible, 
we have to have  
\begin{eqnarray}
 (\fmslash{v} - 1) Q_v (x) | B (v) \rangle  &=& 0  \label{sB0} \\
  i \fmslash{D}^\perp  Q_v (x) | B (v) \rangle  &=& 0   \label{KUL} 
\end{eqnarray}  
Note that (\ref{sB0}) implies that  $Q_v (x) | B (v) \rangle$ only has ``large'' components, while 
(\ref{KUL}) is the relativistic analogon of  (\ref{nrsB}). 

In conclusion, we do not have any proof of this particular limit, nor is the (super)symmetry of the simple quantum mechanical 
model discussed above present in QCD, even if we treat the gluon fields as classical external fields. Likewise, 
the operator $\bar{Q}_v \gamma_5 i \fmslash{D}^\perp  Q_v $ (or alike) cannot evidently be interpreted as a generator 
of a symmetry. 

Nevertheless, motivated by the phenomenological findings we consider this limit as being not an accident. Thus we 
consider the relations (\ref{sB0}) and (\ref{KUL}) as approximate relations in QCD, at least we consider them useful for 
constraining the higher-order matrix elements appearing in the heavy quark expansion. These matrix elements have the generic form 
$$
\langle B(v) |\bar{Q}_v  \Gamma (iD_{\mu_1}) (iD_{\mu_2})  \cdots (iD_{\mu_n}) Q_v | B(v) \rangle  
$$
where $\Gamma$ is an arbitrary Dirac matrix. The NU limit implies, in particular the relation (\ref{KUL}) for the state $Q_v | B(v) \rangle$,  
implies for the matrix elements under consideration 
\begin{equation}  \label{KUL1}
\langle B(v) |\bar{Q}_v  \Gamma (iD_{\mu_1}) (iD_{\mu_2})  \cdots (iD_{\mu_{n-1}}) (i \fmslash{D}_\perp) Q_v | B(v) \rangle  = 0 \, . 
\end{equation} 
We shall exploit this in the following to derive relation between the parameters appearing in the HQE up to $1/m_Q^5$. 

\section{Application to HQE matrix elements} 
\label{HOM} 
At dimension five the parameters are defined as 
\begin{eqnarray} 
2 M_B \mu_\pi^2 &=& - \langle B(v) |\bar{Q}_v (iD_\perp)^2 Q_v | B(v) \rangle \\ 
2 M_B \mu_G^2 &=&  \langle B(v) |\bar{Q}_v (iD_\perp^\mu) (iD_\perp^\mu) i \sigma_{\mu \nu}  Q_v | B(v) \rangle 
\end{eqnarray}  
The relation in the NU limit is obtained from 
\begin{equation}
 \langle B(v) |\bar{Q}_v (i \fmslash{D}_\perp) (i \fmslash{D}_\perp) Q_v | B(v) \rangle  = 0 
\end{equation} 
which immediately yields $\mu_\pi^2 = \mu_G^2$.  

At dimension six the reasoning is the same. The definition of the parameters is   
\begin{eqnarray} 
2 M_B \rho_D^3 &=& - \langle B(v) |\bar{Q}_v (iD_\perp^\mu) (ivD)  (iD_{\perp \mu})  Q_v | B(v) \rangle \\ 
2 M_B \rho_{LS}^3 &=&  \langle B(v) |\bar{Q}_v (iD_\perp^\mu)  (ivD)  (iD_\perp^\mu) i \sigma_{\mu \nu}  Q_v | B(v) \rangle 
\end{eqnarray}  
The relation in the NU limit is obtained from 
\begin{equation}
 \langle B(v) |\bar{Q}_v (i \fmslash{D}_\perp) (ivD) (i \fmslash{D}_\perp) Q_v | B(v) \rangle  = 0 
\end{equation} 
which immediately yields $\rho_D^3 = - \rho_{LS}^3$.   

One of the reasons to reconsider this limit are the most recent fits for these parameters from inclusive semileptonic decays. 
One obtains (taken form the default fit in~\cite{Gambino:2013rza})
\begin{eqnarray}
\mu_\pi^2 &=&  (0.414 \pm 0.078) \, \mbox{GeV}^2  = [ (0.643 \pm 0.061)\, \mbox{GeV} \, ]^2 \\
\mu_G^2 &=&   (0.340 \pm 0.066  ) \, \mbox{GeV}^2 = [ (0.583 \pm 0.057)\, \mbox{GeV} \, ]^2  \\
\rho_D^3 &=& (0.154 \pm 0.045 ) \, \mbox{GeV}^3 = [ (0.536 \pm 0.052)\, \mbox{GeV} \, ]^3 \\
\rho_{LS}^3 &=&  - (0.147 \pm 0.098 ) \, \mbox{GeV}^3 = -  [ (0.528 \pm 0.117)\, \mbox{GeV} \, ]^3 
\end{eqnarray} 
indicating that nature is close to the NU limit. 

At dimension seven we have overall nine parameters, which are defined as~\cite{Heinonen:2014dxa} 
\begin{subequations} \label{eq:defofms}
\begin{align}
2M_B m_1 & = \Braket{B| \bar b_v iD^\perp_\mu iD^\perp_\nu iD^\perp_\rho iD^\perp_\sigma b_v |B} 
\frac{1}{3}(g_\perp^{\mu\nu}g_\perp^{\rho\sigma}+g_\perp^{\mu\rho}g_\perp^{\nu\sigma}+g_\perp^{\mu\sigma}g_\perp^{\nu\rho})\\
2M_B m_2 & = \Braket{B| \bar b_v [iD^\perp_\mu,iv\cdot D][i v \cdot D, iD^\perp_\sigma] b_v |B} g_\perp^{\mu\sigma}\\
2M_B m_3 & = \Braket{B| \bar b_v [iD^\perp_\mu,iD^\perp_\nu][iD^\perp_\rho, iD^\perp_\sigma] b_v |B} g_\perp^{\mu\rho} g_\perp^{\nu\sigma}\\
2M_B m_4 & = \Braket{B| \bar b_v \Big\{ iD^\perp_\mu,\big[iD^\perp_\nu,[iD^\perp_\rho, iD^\perp_\sigma ] \big] \Big\} b_v |B} g_\perp^{\nu\rho} g_\perp^{\mu\sigma}\\
2M_B m_5 & = -\Braket{B| \bar b_v [iD^\perp_\mu,iv\cdot D][i v \cdot D, iD^\perp_\sigma] i\sigma_\perp^{\mu\sigma}b_v |B} \\
2M_B m_6 & = -\Braket{B| \bar b_v [iD^\perp_\mu,iD^\perp_\nu][iD^\perp_\rho, iD^\perp_\sigma] i \sigma_\perp^{\nu\rho} b_v |B}  g_\perp^{\mu\sigma}\\
2M_B m_7 & = -\Braket{B| \bar b_v \Big\{ \{ iD^\perp_\mu,iD^\perp_\nu\}, [iD^\perp_\rho, iD^\perp_\sigma] \Big\} i \sigma_\perp^{\mu\sigma} b_v |B}  
g_\perp^{\nu\rho}\\
2M_B m_8 & = -\Braket{B| \bar b_v \Big\{ \{ iD^\perp_\mu,iD^\perp_\nu\}, [iD^\perp_\rho, iD^\perp_\sigma] \Big\} i \sigma_\perp^{\rho\sigma} b_v |B}  
g_\perp^{\mu\nu}\\
2M_B m_9 & = -\Braket{B| \bar b_v \Big[ iD^\perp_\mu,\big[iD^\perp_\nu,[iD^\perp_\rho, iD^\perp_\sigma ] \big] \Big] i \sigma_\perp^{\rho\mu} b_v |B} 
g_\perp^{\nu\sigma}
\end{align}
\end{subequations} 
Note that we have four spin-singlet operators ($m_1$, ... ,$m_4$) and five spin-triplet operators ($m_5$, ... ,$m_9$). 

Using the NU limit (\ref{KUL}) we obtain relations between the parameters $m_i$, $i=1,..,9$, which are obtained by contracting
the remaining indices with the possible vectors and Dirac matrices. For example, we have 
\begin{equation}
\langle B(v) |\bar{Q}_v  (i \fmslash{D}_\perp)  (ivD)^2 (i \fmslash{D}_\perp) Q_v | B(v) \rangle  = 0
\end{equation} 
from which we directly obtain $m_2 + m_5 = 0$. 

At dimension seven it turns out that we have overall  five independent relations; 
we may use these to express the spin-triplet operators
in terms of the spin-singlet ones. These relations are 
\begin{subequations} \label{mi} 
\begin{align} 
m_5 &= - m_2  \vphantom{\frac{1}{3}} \\ 
m_6 &= m_1 + m_3 + \frac{1}{3} m_4  \\
m_7 &= - 2 m_1 - \frac{2}{3} m_4 \\
m_8 &= - 8 m_1 - 4 m_3 - \frac{8}{3} m_4 \\
m_9 &= - m_4 \vphantom{\frac{1}{3}} 
\end{align}
\end{subequations}

At dimension eight we have overall eighteen parameters,  out of which seven ($r_1$, ..., $r_7$) are spin-singlet operators. Note that 
the number of independent spin-triplet operators grows faster than the number of spin-singlet ones, since the two 
indices of the spin matrix allow for more independent contractions. We define these matrix elements as~\cite{Heinonen:2014dxa} 
\begin{subequations}
\begin{align}
2M_B r_1 &= \Braket{ B | \bar b_v \,i  D_\mu^\perp\, (i v \cdot D)^3\, i  D^\mu_\perp \, b_v | B } \\
2M_B r_2 &= \Braket{ B | \bar b_v \,i  D_\mu^\perp\, (i v \cdot D)\, i  D^\mu_\perp\, i  D_\nu^\perp\, i  D^\nu_\perp \, b_v | B } \\
2M_B r_3 &= \Braket{ B | \bar b_v \,i  D_\mu^\perp\, (i v \cdot D)\, i  D_\nu^\perp\, i D_\perp^\mu\, i  D_\perp^\nu \, b_v | B } \\  
2M_B r_4 &= \Braket{ B | \bar b_v \,i  D_\mu^\perp\, (i v \cdot D)\, i  D_\nu^\perp\, i D^\nu_\perp\, i  D^\mu_\perp \, b_v | B } \\  
2M_B r_5 &= \Braket{ B | \bar b_v \,i  D_\mu^\perp\, i  D^\mu_\perp\,(i v \cdot D)\,  i D_\nu^\perp\, i  D^\nu_\perp \, b_v | B } \\  
2M_B r_6 &= \Braket{ B | \bar b_v \,i  D_\mu^\perp\, i  D_\nu^\perp\, (i v \cdot D)\, i D^\nu_\perp\, i  D^\mu_\perp \, b_v | B } \\  
2M_B r_7 &= \Braket{ B | \bar b_v \,i  D_\mu^\perp\, i  D_\nu^\perp\, (i v \cdot D)\, i D^\mu_\perp\, i  D^\nu_\perp \, b_v | B } \\  
2M_B r_{8} &= - \Braket{ B | \bar b_v \,i   D_\alpha^\perp \, (i v \cdot D)^3\, i   D_\beta^\perp \, i \sigma_\perp^{\alpha \beta }\,b_v | B } \\  
2M_B r_{9} &= - \Braket{ B | \bar b_v \,i   D_\alpha^\perp \, (i v \cdot D)\, i   D_\beta^\perp \, i   D_\mu^\perp\, i   D^\mu_\perp \,i \sigma_\perp^{\alpha \beta }\, b_v | B } \\  
2M_B r_{10} &= - \Braket{ B | \bar b_v \,i   D_\mu^\perp\, (i v \cdot D)\, i D^\mu_\perp\, i   D_\alpha^\perp \, i   D_\beta^\perp  \,i \sigma_\perp^{\alpha \beta }\, b_v | B } \\
2M_B r_{11} &= - \Braket{ B | \bar b_v \,i   D_\mu^\perp\, (i v \cdot D)\, i   D_\alpha^\perp \, i   D^\mu_\perp\, i   D_\beta^\perp  \,i \sigma_\perp^{\alpha \beta }\, b_v | B } \\  
2M_B r_{12} &= -\Braket{ B | \bar b_v \,i   D_\alpha^\perp \, (i v \cdot D)\, i D_\mu^\perp\, i   D_\beta^\perp \, i   D^\mu_\perp \,i \sigma_\perp^{\alpha \beta }\, b_v | B } \\
2M_B r_{13} &= - \Braket{ B | \bar b_v \,i   D_\mu^\perp\, (i v \cdot D)\, i   D_\alpha^\perp \, i   D_\beta^\perp \, i   D^\mu_\perp \,i \sigma_\perp^{\alpha \beta }\, b_v | B } \\  
2M_B r_{14} &= - \Braket{ B | \bar b_v \,i   D_\alpha^\perp \, (i v \cdot D)\, i D_\mu^\perp\, i   D^\mu_\perp\, i   D_\beta^\perp  \, i \sigma_\perp^{\alpha \beta }\, b_v | B } \\
2M_B r_{15} &= - \Braket{ B | \bar b_v \,i   D_\alpha^\perp \, i   D_\beta^\perp \, (i v \cdot
D)\, i   D_\mu^\perp\, i   D^\mu_\perp \, i \sigma^{\alpha \beta }\, b_v | B } \\  
2M_B r_{16} &= - \Braket{ B | \bar b_v \,i   D_\mu^\perp\, i   D_\alpha^\perp \, (i v \cdot
D)\, i   D_\beta^\perp \, i   D^\mu_\perp \, i \sigma_\perp^{\alpha \beta }\, b_v | B } \\  
2M_B r_{17} &= - \Braket{ B | \bar b_v \,i   D_\alpha^\perp \, i   D_\mu^\perp\, (i v \cdot
D)\, i   D^\mu_\perp\, i   D_\beta^\perp  \, i \sigma_\perp^{\alpha \beta }\, b_v | B } \\  
2M_B r_{18} &= -\Braket{ B | \bar b_v \,i   D_\mu^\perp\, i   D_\alpha^\perp \, (i v \cdot D)\, i   D^\mu_\perp\, i   D_\beta^\perp  \, i \sigma_\perp^{\alpha \beta }\, b_v | B } \,.
\end{align}
\end{subequations}

Using the trace formulae shonw in the appendix, we obtain in total nine independent relation between the matrix elements in the 
NU limit (\ref{KUL}). Since the number of spin triplet operators is now larger than the number of independent relations, we cannot express 
the spin triplet operators in terms of the singlet ones any more. One way of writing the nine relations is 
\begin{subequations} \label{ri} 
\begin{align} 
r_1 + r_8 &= 0 \vphantom{\frac{1}{1}}  \\ 
r_{10} +r_2 & = 0 \vphantom{\frac{1}{1}} \\
r_{11} + r_{3} & = 0 \vphantom{\frac{1}{1}}  \\
r_{14} + r_4 &= 0  \vphantom{\frac{1}{1}}  \\
-\frac{2 r_{16}}{5} + \frac{r_{17}}{5} + \frac{2 r_{18}}{5} + r_5 &= 0  \\ 
r_{17}+r_6 &= 0 \vphantom{\frac{1}{1}}  \\
r_{18} + r_7 &= 0 \vphantom{\frac{1}{1}} \\
- r_{12} + r_{13} + r_9 &= 0 \vphantom{\frac{1}{1}}  \\ 
r_{15} + \frac{4 r_{16}}{5} - \frac{2 r_{17}}{5} + \frac{r_{18}}{5} & = 0  
\end{align}
\end{subequations}
In many cases these are again relations between pairs of spin-singlet and -triplett operators, which has a similar structure. 

\section{Discussion and Conclusion}  
Assuming that real QCD is close to the NU limit, we can significantly reduce the number of independent 
parameters appearing in higher orders of the HQE, roughly by a factor of two. At dimension seven the number 
of independent parameters is reduced from nine to four, while at dimension eight we find a reduction from 
eighteen to nine independent parameters. This nevertheless leaves 
still a large number of parameters, which need to be determined either form theoretical input of from data. 

One may compare the above results with the estimates for the matrix elements obtained from ``lowest state saturation 
assumption'' (LSSA), which corresponds to naive factorization. While this approach suffers from the same problems as 
the naive factorization in non-leptonic decays, it is up to now the only way to obtain numerical values for the parameters 
$m_i$ and $r_i$ beyond a pure guess. 

It has been pointed out aready in \cite{Heinonen:2014dxa} that in LSSA the combination $\rho_D^3 + \rho_{LS}^2$ 
receives only contribution of the $j=1/2$ positive parity doublet of the first excited states. The BPS limit in this case 
implies the complete absence of this contribution; in LSSA this contribution is proportional to $\mu_\pi^2 - \mu_G^2$ 
and hence LSSA for the dimension five and six matrix elements is compatible with the BPS limit. 

The LSSA relates the higher-order terms to the lower order ones, comparing the results  given in LSSA in 
\cite{Heinonen:2014dxa} we find the relations (\ref{mi}) and (\ref{ri}) are indeed satisfied once the leading 
relation  $\mu_\pi^2 - \mu_G^2 = 0$ is assumed. Thus we conclude that LSSA is  compatible with the BPS limit; 
this fact is also reflected by the numerical values presented in \cite{Heinonen:2014dxa}.  

The BPS limit has also some other applications which have beed discussed in the orignal paper \cite{Uraltsev:2003ye} 
and some more applications can  be found. In particular, using (\ref{KUL}) one may apply this also to e.g. nonlocal matrix 
elements, which would result in relations between shape functions and light cone distributions with different spin 
configurations. 

As it is now, the application of the NU limit relies only on some phenomenological evidence; if more evidence can be 
accumulated, it would be interesting to understand the dynamical origin of this limit in QCD. 
Eventually it could become a useful tool  including a way to compute corrections to this limit. 

\section*{Acknowledgements} 
T.M. thanks Thorsten Feldmann for clarifying discussions. 
This work was supported by the DFG research unit DFG FOR 1873 ``Quark Flavour Phsyics and Effective Field Theories'' 
\appendix

\section{Trace Formulae}
To avoid notational clutter the spinor indices are suppressed and projection matrices omitted. In this notation we write
\begin{align}
\braket{B|\bar b_v ( ... ) b_v|B} = X_{singlet} + X_{triplet} (-i \sigma^\perp)
\end{align}
which is really
\begin{align}
\braket{B|\bar b_{v,\alpha}  ...  b_{v,\beta}|B} = X_{singlet} (P_+)_{\beta \alpha} + X_{triplet} (-i P_+ \sigma^\perp P_+)_{\beta \alpha}.
\end{align}
We will give for each dimension only the leading term in the $1/m_b$ expansion.  

\subsection{For dimension five matrix elements} 
We have
\begin{align}
\frac{1}{2 M_B}\braket{B|\bar b_v iD_\mu iD_\nu  b|B} = -\frac{\mu_\pi^2}{6} g^\perp_{\mu\nu}  - \frac{\mu_g^2}{12} (- i \sigma^{\perp}_{\mu\nu}).
\end{align}

\subsection{For dimension six matrix elements} 
We have
\begin{align}
\frac{1}{2 M_B}\braket{B|\bar b_v iD_\mu iD_\nu iD_\rho  b|B} = \frac{\rho_D^3}{6} v_{\nu} g^\perp_{\mu\rho}  - \frac{\rho_{LS}^3}{12} v_{\nu} (- i \sigma^{\perp}_{\mu\rho}).
\end{align}

\subsection{For dimension seven matrix elements} 
Obviously, the choice of basis functions is not unique. Furthermore, there are more independent Lorentz structures for the triplet terms, however only certain combinations yield Hermitian operators in the matrix element.  One choice is
\begin{align}
\frac{1}{2 M_B}\braket{B|\bar b_v iD_\mu iD_\nu iD_\rho iD_\sigma  b|B} & = 
c_1 \left[ g^\perp_{\mu\nu} g^\perp_{\rho\sigma} + g^\perp_{\mu\rho} g^\perp_{\nu\sigma} + g^\perp_{\mu\sigma} g^\perp_{\rho\nu} \right]  
 + c_2 \, v_\nu v_\rho g^\perp_{\mu\sigma} \nonumber \\
& \quad + c_3 \left[ g^\perp_{\mu\nu} g^\perp_{\rho\sigma} - g^\perp_{\mu\sigma} g^\perp_{\rho\nu} \right]  
+ c_4 \left[ 2 g^\perp_{\mu\nu} g^\perp_{\rho\sigma} - g^\perp_{\mu\rho} g^\perp_{\nu\sigma} - g^\perp_{\mu\sigma} g^\perp_{\rho\nu} \right]  \nonumber \\
& \quad + c_5 \, v_\nu v_\rho (-i \sigma^\perp_{\mu\sigma}) \\
& \quad + c_6 \left[ g^\perp_{\mu\nu} (-i \sigma^\perp_{\rho\sigma}) + g^\perp_{\rho\sigma} (-i \sigma^\perp_{\mu\nu}) \right]
+ c_7  \, g^\perp_{\mu\sigma} (-i \sigma^\perp_{\nu\rho})  \nonumber \\
& \quad + c_8 \left[ \epsilon_{\alpha\mu\kappa\lambda}v^\alpha\epsilon_{\beta\nu\rho\sigma}v^\beta  +\epsilon_{\alpha\mu\nu\rho}v^\alpha\epsilon_{\beta\sigma\kappa\lambda}v^\beta \right] (-i \sigma_\perp^{\kappa\lambda}) \nonumber \\
& \quad + c_9 \left[ \epsilon_{\alpha\mu\kappa\lambda}v^\alpha\epsilon_{\beta\nu\rho\sigma}v^\beta  -\epsilon_{\alpha\mu\rho\sigma}v^\alpha\epsilon_{\beta\nu\kappa\lambda}v^\beta \right] (-i \sigma_\perp^{\kappa\lambda}). \nonumber
\end{align}
With this choice the coefficients for the singlet terms and the triplet term which contains $v$ are simple
\begin{subequations}
\begin{align}
 c_1 & =  \frac{m_1}{30}, \\
 c_2 & =  \frac{m_2}{12},  \\
 c_3 & =  \frac{m_3}{24}, \\
 c_4 & =  \frac{m_4}{72}  ,\\
 c_5 & =  -\frac{m_5}{12} , 
 \intertext{while the remaining triplet coefficients are given by}
\begin{pmatrix}
c_6 \\ c_7 \\ c_8 \\ c_9 
\end{pmatrix} 
& = 
\frac{1}{960}
\begin{pmatrix}
	-24 & -24  & -6 & 16 \\
	-48 & -16  & -8 & 32 \\
	-12 & -10  & 0 & 8 \\
	16 & 8  & -1 & -4 \\
\end{pmatrix} 
\begin{pmatrix}
m_6 \\ m_7 \\ m_8 \\ m_9 
\end{pmatrix}.
\end{align}
\end{subequations}

\subsection{For dimension eight matrix elements} 
Again, as in the dimension seven case the choice of basis functions is not unique. One choice is
\begin{align}
& \frac{1}{2 M_B}\braket{B|\bar b_v iD_\mu iD_\nu iD_\rho iD_\sigma iD_\lambda  b|B} 
	\nonumber \\
& \qquad 
	= d_1 \, v_\nu v_\rho v_\sigma g^\perp_{\mu\lambda} 
	+ d_2 \, v_\nu g^\perp_{\mu\rho}  g^\perp_{\sigma\lambda} 
	+ d_3 \, v_\nu g^\perp_{\mu\sigma}  g^\perp_{\rho\lambda} 
	+ d_4 \, v_\nu g^\perp_{\mu\lambda}  g^\perp_{\rho\sigma} 
	\nonumber \\
& \qquad \qquad 
	+ d_5 \, v_\rho g^\perp_{\mu\nu}  g^\perp_{\sigma\lambda} 
	+ d_6 \, v_\rho g^\perp_{\mu\lambda}  g^\perp_{\nu\sigma} 
	+ d_7 \, v_\rho g^\perp_{\mu\sigma}  g^\perp_{\nu\lambda} 
	\nonumber  \\
& \qquad \qquad 
	+ d_8 \, v_\nu v_\rho v_\sigma (-i \sigma^\perp_{\mu\lambda})
	+ d_9 \, v_\nu g^\perp_{\sigma\lambda}  (-i \sigma^\perp_{\mu\rho})
	+ d_{10} \, v_\nu g^\perp_{\mu\rho}  (-i \sigma^\perp_{\sigma\lambda})
	\\
& \qquad \qquad 
	+ d_{11} \, v_\nu  g^\perp_{\mu\sigma}  (-i \sigma^\perp_{\rho\lambda})
	+ d_{12} \, v_\nu g^\perp_{\rho\lambda}  (-i \sigma^\perp_{\mu\sigma})
	+ d_{13} \, v_\nu  g^\perp_{\mu\lambda}  (-i \sigma^\perp_{\rho\sigma})
	+ d_{14} \, v_\nu g^\perp_{\rho\sigma}  (-i \sigma^\perp_{\mu\lambda})
	\nonumber  \\
& \qquad \qquad 
	+ d_{15} \, v_\rho g^\perp_{\sigma\lambda}  (-i \sigma^\perp_{\mu\nu})
	+ d_{16} \, v_\rho g^\perp_{\mu\lambda}  (-i \sigma^\perp_{\nu\sigma})
	+ d_{17} \, v_\rho g^\perp_{\nu\sigma}  (-i \sigma^\perp_{\mu\lambda})
	+ d_{18} \, v_\rho g^\perp_{\mu\sigma}  (-i \sigma^\perp_{\nu\lambda})
	\nonumber 
\end{align}
With this choice the coefficients for the singlet terms are given by
\begin{subequations}
\begin{align}
d_1 & =  \frac{r_1}{6}, \\
\begin{pmatrix}
d_2 \\ d_3 \\ d_4 
\end{pmatrix} 
& = 
\frac{1}{60}\begin{pmatrix}
4 & -1 & -1 \\
-1 & 4 & -1 \\
-1 & -1 & 4 
\end{pmatrix} 
\begin{pmatrix}
r_2 \\ r_3 \\ r_4 
\end{pmatrix} ,
	\\
\begin{pmatrix}
d_5 \\ d_6 \\ d_7 
\end{pmatrix} 
& = 
\frac{1}{60}\begin{pmatrix}
4 & -1 & -1 \\
-1 & 4 & -1 \\
-1 & -1 & 4 
\end{pmatrix} 
\begin{pmatrix}
r_5 \\ r_6 \\ r_7 
\end{pmatrix}, 
 \intertext{while the remaining triplet coefficients are}
d_8 & = -\frac{r_8}{6}, \\
\begin{pmatrix}
d_9 \\ d_{10} \\ d_{11} \\ d_{12} \\ d_{13} \\ d_{14} 
\end{pmatrix} 
& = 
\frac{1}{30}\begin{pmatrix}
	-3 & 0  & -1 & 1 & -1 & 1 \\ 
	0 & -3  & 1 & -1 & -1 & 1 \\ 
	-1 & 1  & -3 & 0 & 1 & 1 \\ 
	1 & -1  & 0 & -3 & 1 & 1 \\ 
	-1 & -1  & 1 & 1 & -3 & 0 \\ 
	1 & 1  & 1 & 1 & 0 & -3 
\end{pmatrix} 
\begin{pmatrix}
r_9 \\ r_{10} \\ r_{11} \\ r_{12} \\ r_{13} \\ r_{14}
\end{pmatrix}, 
	\\
\begin{pmatrix}
 d_{15} \\ d_{16} \\ d_{17} \\ d_{18} 
\end{pmatrix} 
& = 
\frac{1}{240}\begin{pmatrix}
	-21 & -4  & 10 & -9 \\
	-4 & -16  & 0 & 4 \\
	10 & 0  & -40 & 10 \\
	-9 & 4  & 10 & -21 
\end{pmatrix} 
\begin{pmatrix}
 r_{15} \\ r_{16} \\ r_{17} \\ r_{18}
\end{pmatrix}.
\end{align}
\end{subequations}

\end{document}